\documentclass[twocolumn,prl,aps,showpacs]{revtex4}
\usepackage{amssymb}

\usepackage{amsmath}
\usepackage{graphicx}
\usepackage{dcolumn}
\usepackage{bm}

\begin{document}

\preprint{APS/123-QED}
\title{Berry Phase Effect on Exciton Transport and Bose Einstein Condensate}
\author{Wang Yao}
\thanks{wangyao@physics.utexas.edu}
\author{Qian Niu}
\affiliation{Department of Physics, The University of Texas, Austin,
Texas 78712}
\date{\today}

\begin{abstract}
With exciton lifetime much extended in semiconductor quantum-well
structures, exciton transport and Bose-Einstein condensation become
a focus of research in recent years. We reveal a momentum-space
gauge field in the exciton center-of-mass dynamics due to Berry
phase effects. We predict spin-dependent topological transport of
the excitons analogous to the anomalous Hall and Nernst effects for
electrons. We also predict spin-dependent circulation of a trapped
exciton gas and instability in an exciton condensate in favor of
vortex formation.
\end{abstract}

 \maketitle


In semiconductors, an electron can be excited from the valance band
into the conduction band by absorbing a photon. Exciton is a bound
state of the extra electron in the conduction band and the hole left
behind in the valance band. Exciton plays a crucial role in
semiconductor optics, e.g. as a source of nonlinear optical
effects~\cite{Chemla_ReviewSemiconductorOptics}. In a bulk
semiconductor, exciton lifetime is quite short because of
recombination annihilation of the electron and hole accompanied by
emission of a photon. By confining electrons and holes separately in
two coupled quantum
wells~\cite{Alexandrou_indirectExciton,Butov_exciton93}, the
resulting indirect exciton can have a lifetime greatly extended (up
to 10 $\mu$s~\cite{Snoke_Exciton_diffusion}), opening up an entirely
new realm of exciton physics. It is of great interest to study
transport of a metastable exciton gas, which may bring in new
functionalities of optoelectronics~\cite{ExcitonicCircuit}.
Moreover, recent experiments suggest that realization of the
Bose-Einstein condensation (BEC) of excitons may finally become in
reach after decades of prediction~\cite{Butov_BEC,Snoke_ring,
Butov_coherenceLength,Butov_repulsive}.

In this Letter, we study the center-of-mass dynamics of excitons,
finding a momentum-space gauge field due to Berry phase in the basis
Bloch functions of the quantum wells as well as in the wavefunction
describing the relative motion of the electron-hole
pair~\cite{Berry_BerryPhase}. The gauge field can lead to
spin-dependent exciton transport much like the anomalous Hall and
Nernst effects for
electrons~\cite{SHE,Chang_BerryPhase,TKNN,with_TKNN,Yao_OSHE,Xiao_ANE}.
As the spin of an exciton is correlated with the polarization of the
emitted photon from exciton annihilation, these anomalous transport
phenomena may be directly observed from polarization-resolved
exciton luminescence. The gauge field can also induce circulation in
a trapped gas of excitons and a spontaneous vortex formation in an
exciton BEC.

{\it Gauge Structure in Exciton Wavefunction.}---In a homogeneous
system, the exciton energy-momentum eigenstate is parameterized by
the center-of-mass wavevector $\mathbf{q}$ and the quantum number
$n$ for each hydrogen-like orbit of the relative motion. The
wavefunction can be generally written as $\Psi^{\rm
ex}_{n,\mathbf{q}} = e^{i \mathbf{q}\cdot \mathbf{R}} U^{\rm
ex}_{n,\mathbf{q}}$ with $\mathbf{R}$ being the center-of-mass
coordinate. Like the Bloch function, the exciton wavefunction is
decomposed into a plane-wave envelope function for center-of-mass
motion and an `internal' structure $U^{\rm ex}_{n,\mathbf{q}} =
\sum_{\mathbf{k}}F_n \left( \mathbf{k,q}\right) e^{i \mathbf{k}\cdot
\mathbf{r}} u_{e, \mathbf{k}+\frac{m_{e}}{M}\mathbf{q}}
u_{h,-\mathbf{k}+\frac{ m_{h}}{M}\mathbf{q}} $ where $\mathbf{k}$
and $\mathbf{r}$ are respectively the wavevector and coordinate for
the relative motion. $u_{e}$ and $u_{h}$ are the periodic part of
the electron and hole Bloch function, and $\sum_{\mathbf{k}}F_n
\left( \mathbf{k,q}\right) e^{i \mathbf{k}\cdot \mathbf{r}}$ gives
the envelope function of the relative motion which may depend on
$\mathbf{q}$ in general. $M \equiv m_e+m_h$ is the exciton mass.
Similar to that of the Bloch electrons~\cite{Chang_BerryPhase}, the
gauge structure of exciton lies in the dependence of the `internal'
structure $U^{\rm ex}_{n,\mathbf{q}}$ on the dynamical parameter
$\mathbf{q}$~\cite{Berry_BerryPhase}. The gauge potential is defined
as $\mathcal{A}^{\rm ex}_{\mu} \equiv i \left\langle U^{\rm ex}
|\partial_{q_{\mu}} | U^{\rm ex}\right\rangle$, and the gauge field
is then $\bm{\Omega}^{\rm ex} \equiv \bm{\nabla}_{\bf{q}} \times
\bm{\mathcal{A}}^{\rm ex}$. This gauge field, known as the Berry
curvature, is analogous to a `magnetic' field in the crystal
momentum space. Its integral over a $q$-space area yields the Berry
phase of an exciton state adiabatically going around the boundary of
the area, which is similar to the relationship between a magnetic
field and the Arharonov-Bohm phase.

\begin{figure}[t]
\includegraphics[width=8 cm]{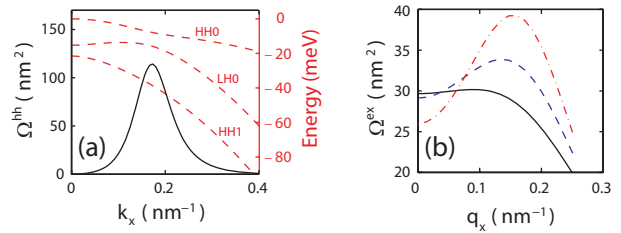}
\caption{Gauge structure in a $10$nm quantum well. (a) Dashed curves
denote the three highest valence subbands. Berry curvature (solid
curve) in heavy-hole subband HH0 is most pronounced where HH0
anti-cross with the light-hole subband LH0. In the calculation, we
assume a valence barrier height of $140$ meV and the same Luttinger
parameters ($\gamma_1=6.85$, $\gamma_2=2.1$, and $\gamma_3=2.9$) for
the quantum well layer and the barrier. (b) Berry curvature as a
function of center-of-mass wavevector for $1s$ heavy-hole exciton.
We assume the in-plane exciton Bohr radius $a_B=$8, 10, and 12~nm
for solid, dashed and dash-dotted curves respectively.}
\label{gaugefield}
\end{figure}

For 2D exciton, the Berry curvature is always normal to the plane
with magnitude given by,
\begin{eqnarray}
\Omega^{\rm ex}\left( \mathbf{q}\right) &=& \left( \frac{m_{h}
}{M}\right) ^{2}\sum_{\mathbf{k}}\left\vert F \right\vert
^{2}\Omega^{h}\left( -\mathbf{k+}\frac{m_{h}}{M}\mathbf{\mathbf{
q}}\right) \label{curv} \\
&+& i\sum_{\mathbf{k}}\left(
\partial_{q_x} F^{\ast }
\partial_{q_y} F-\partial_{q_y}
F^{\ast } \partial_{q_x} F \right) \notag \\
&+&\frac{m_{h}}{M}\sum_{\mathbf{k}} \left[ \partial_{q_x} \left\vert
F \right\vert ^{2}\mathcal{A}^h_{y}\left( -
\mathbf{k+}\frac{m_{h}}{M}\mathbf{\mathbf{q}}\right) - x
\leftrightarrow y \right] \notag
\end{eqnarray}
where $\mathcal{A}^h_{\mu} (\mathbf{k}) \equiv i \left\langle
u_{h,\mathbf{k}} |\partial_{k_{\mu}} | u_{h,\mathbf{k}}\right\rangle
$ and $\bm{\Omega}^h(\mathbf{k}) \equiv \bm{\nabla}_{\mathbf{k}}
\times \bm{\mathcal{A}}^h (\mathbf{k})$ are respectively the gauge
potential and gauge field in the hole band. The Berry curvature of
the exciton thus has three parts. On the right hand side of
Eq.~(\ref{curv}), the {\bf first} term  is the inheritance of Berry
curvatures from the parent Bloch bands~\cite{curvature_conduction}.
In quantum well, heavy-light hole mixing at finite in-plane
wavevector leads to pronounced Berry curvature distributions in
these subbands~\cite{Yao_OSHE}, and we expect this to be the
dominant contribution to the exciton Berry curvature
[Fig.~(\ref{gaugefield})]. This contribution is {\it spin dependent}
as $\Omega^{h}$ changes sign when hole spin flips. The {\bf second}
term is due to the entanglement of relative motion in $n$th orbit
with the center-of-mass motion. This contribution is spin
independent in general, but can have opposite values at $\bf{q}$ and
$-\bf{q}$, analogous to the valley dependent Berry curvature in
graphene~\cite{valley_curvature}. It must vanish when the system has
both time reversal symmetry and $180^{\circ}$ rotation symmetry in
the 2D plane~\cite{valley_curvature}. The {\bf third} one is the
cross terms of the first two. Its spin dependence comes from the
gauge potential $\mathcal{A}^h_{\mu}$.

{\it Semiclassical Equation of Motion.}---From now on, we focus on
spatially indirect excitons with long radiative lifetime. To see the
Berry phase effect, we first establish the semiclassical theory for
wavepacket dynamics of excitons subject to perturbations varying
slowly in space and time. Consider the following exciton wavepacket
centered at center-of-mass coordinate $\mathbf{R}_c$: $|X\rangle =
\int [d\mathbf{q}] w(\mathbf{q}) \left\vert \Psi^{\rm
ex}_{\mathbf{q}}\right\rangle$. $w(\mathbf{q})$ is a function
localized around $\mathbf{q}_c$ with a width much smaller than the
inverse of the in-plane exciton Bohr radius $a_B$, and $\int
[d\mathbf{q}]$ stands for $\int d\mathbf{q}(2 \pi)^{-2}$ here and
hereafter. The equation of motion for this wavepacket can be derived
from the effective Lagrangian~\cite{Chang_BerryPhase},
\begin{subequations}
\begin{eqnarray}
\bf {\dot{R}}_c &=& \frac{\partial
\mathcal{E}(\mathbf{q}_c,\mathbf{R}_c)}{\hbar \partial \mathbf{q}_c}
- \bf{\dot{q}}_c \times
\bm{\Omega}^{\rm ex}, \label{eom_a}\\
\hbar \bf{\dot{q}}_c &=& -\frac{\partial
\mathcal{E}(\mathbf{q}_c,\mathbf{R}_c)}{\partial \mathbf{R}_c} -
\bf{\dot{R}}_c \times \bm{\mathcal {D}} \label{eom_b}.
\end{eqnarray}
\label{eom}
\end{subequations}
$\mathcal{E}(\mathbf{q}_c,\mathbf{R}_c)$ is the semiclassical energy
of the exciton wavepacket with center-of-mass wavevector
$\mathbf{q}_c$ and coordinate $\mathbf{R}_c$. To leading order, it
is of a factorized form $\mathcal{E}(\mathbf{q}_c,\mathbf{R}_c)
\equiv \mathcal{E}_0(\mathbf{q}_c) + V(\mathbf{R}_c)$.
$\mathcal{E}_0(\mathbf{q}_c)$ is the unperturbed exciton dispersion
in the homogeneous quantum well. $V$ is the potential energy from
external perturbations, so that the exciton center-of-mass motion is
subjected to a mechanical force $\bm{\mathcal{C}} \equiv -
\bm{\nabla} V$. In an electrostatic potential $\phi$,
$\mathcal{C}_{x,y} \equiv ed
\partial_{x,y} \frac{\partial \phi}{\partial z}$ with $d$ being
the separation between the electron and hole
layers~\cite{Butov_trap}. Thus, the intrinsic dipole moment of
indirect exciton allows its transport to be controlled by the
electric field gradient, which forms the basis of electrically gated
excitonic circuits~\cite{ExcitonicCircuit}. The dipole moment also
allows a {\it real-space} `magnetic' field $\bm{\mathcal{D}} \equiv
ed \frac{\partial B_z}{\partial z} \hat{\bm z}$ from the gradient of
the external magnetic field $\bm{B}(\mathbf{r})$. In conjugation,
the Berry curvature $\bm{\Omega}^{\rm ex}$ plays the role of a {\it
momentum-space} `magnetic' field which gives rise to an anomalous
contribution to the velocity~\cite{Chang_BerryPhase,TKNN,with_TKNN}.
For $1s$ heavy-hole excitons being investigated for transport and
condensation phenomena~\cite{Butov_BEC,Snoke_ring,
Butov_coherenceLength,Butov_repulsive,Snoke_Exciton_diffusion}, the
Berry curvature distribution is opposite between the two `bright'
exciton branches $e^{\dagger}_+ h^{\dagger}_-|G\rangle$ and
$e^{\dagger}_- h^{\dagger}_+|G\rangle$, and between the two `dark'
exciton branches $e^{\dagger}_+ h^{\dagger}_+|G\rangle$ and
$e^{\dagger}_- h^{\dagger}_-|G\rangle$~\cite{notation}. The exciton
center-of-mass motion thus acquires a spin-dependent ingredient.

It is interesting to make a comparison with the spin dependent
center-of-mass dynamics recently observed on
exciton-polaritons~\cite{polaritonSHE}, which results from the
momentum dependent polarization splitting for spatially direct
excitons. Such mechanism is unimportant for indirect excitons where
polarization splitting is negligible as the wavefunction overlap
between the electron and hole component is
small~\cite{Littlewood_exciton_BEC_review}. In addition to the
heavy-light hole mixing mechanism, spin dependent center-of-mass
motion of indirect exciton can also arise from Rashba spin-orbit
coupling in the parent Bloch bands~\cite{Li_excitonSHE}.

{\it Spin Hall Effect and Spin Nernst Effect.}---The above
semiclassical equation of motion is the basis for calculation of
exciton transport currents driven by mechanical and statistical
forces. The exciton current density can be defined as $\bm{J} \equiv
\int [d \mathbf{q}_c] f(\mathbf{q}_c,\mathbf{R}_c) \bm{\dot{ R}}_c$
where $f$ is the distribution function. In an electrically
controlled excitonic circuit~\cite{ExcitonicCircuit}, the flow of
excitons is driven by the `electric-like' force $\bm{\mathcal C}$
from patterned electrodes~\cite{Butov_trap}. We immediately find
that the anomalous velocity term in Eq.~(\ref{eom_a}) contributes a
spin-dependent exciton Hall current: $\bm{J}=\bm{\mathcal C} \times
\frac{1}{\hbar}\int [d \mathbf{q}_c ] f(\mathbf{q}_c)
\bm{\Omega}^{\rm ex}(\mathbf{q}_c)$~\cite{with_TKNN}. Note that the
above Hall current comes mainly from the equilibrium part of carrier
distribution and hence is referred as the {\it intrinsic}
contribution, as opposed to the {\it extrinsic} contribution from
the non-equilibrium part of distribution by phonon scattering or
disorder scattering~\cite{SHE,AHE_phonon}. Intrinsic contribution
dominates the anomalous Hall effect when phonon scattering is the
major cause for exciton momentum relaxation~\cite{AHE_phonon}. In
disordered systems at low temperatures, the relative importance of
intrinsic and extrinsic contribution is not fully understood, and
exciton transport can be further complicated by quantum interference
effects~\cite{interference}, which is, however,  beyond the scope of
this work.


In most current experiments, hot indirect excitons are generated at
laser excitation spot, and the exciton temperature decreases by
phonon emissions upon diffusion to remote trap
regions~\cite{Butov_BEC,Snoke_ring,
Butov_coherenceLength,Butov_repulsive}. `Thermoelectric' responses
to the statistical forces of temperature gradient and chemical
potential gradient is thus of direct relevance. From the basis of
the Einstein relation, it is suggested that an exciton spin Hall
current can be induced by a chemical potential gradient.
Furthermore, the Mott relation for the `electrical' conductivity and
`thermoelectric' conductivity suggests the {\it spin Nernst effect},
i.e. spin Hall current driven by a temperature gradient. Both
relations are proven to hold for Berry phase supported topological
transport currents of Bloch electrons~\cite{Xiao_ANE}, and the
conclusion is straightforwardly generalized to excitons of Boson
statistics. Specifically, we find the spin-dependent exciton Hall
current in the presence of chemical potential gradient and
temperature gradient: $\bm{J}=-\bm{\nabla}\mu \times \frac{1}{\hbar}
\int [d \mathbf{q}_c] f \bm{\Omega}^{\rm ex} -\frac{\bm{\nabla}T}{T}
\times \frac{1}{\hbar} \int [d \mathbf{q}_c] \bm{\Omega}^{\rm ex}
\left[(\mathcal{E}- \mu) f - k_B T \log(1 - e^{-\beta(\mathcal{E}-
\mu)}) \right]$.

One can extract the Hall conductivity $\sigma$ and the Nernst
conductivity $\alpha$ defined by $J_x=\sigma (\mathcal{C}_y
-\nabla_y \mu)+\alpha (-k_B\nabla_y T)$. The exciton Berry curvature
from heavy-light hole mixing is most pronounced in a momentum space
region centered around $\bf{q}=0$ [Fig.\ref{gaugefield}(b)].
Therefore, we may approximate $\Omega^{\rm ex} \propto
\theta(\mathcal{E}_s-\mathcal{E})$ where typically $\mathcal{E}_s
\sim$~meV. Assuming $k_BT \gg \mathcal{E}_s$, we find
\begin{equation}
\sigma=\frac{1}{h}\frac{\Phi}{2\pi}\frac{\zeta}{1-\zeta}, ~~
\alpha=\frac{1}{h}\frac{\Phi}{2\pi} \frac{\zeta \ln \zeta +
(1-\zeta) \ln (1-\zeta)}{\zeta-1}. \notag
\end{equation}
$\zeta \equiv \exp(\mu/k_B T)$ is the {\it fugacity} of the exciton
gas and $\Phi\equiv \int d \mathbf{q} \Omega^{\rm ex}(\mathbf{q}) $
is the total flux of the Berry curvature. For typical GaAs quantum
wells, we find $\Phi\sim\pi$.

\begin{figure}[t]
\includegraphics[width=8cm,bb=22 282 593 542]{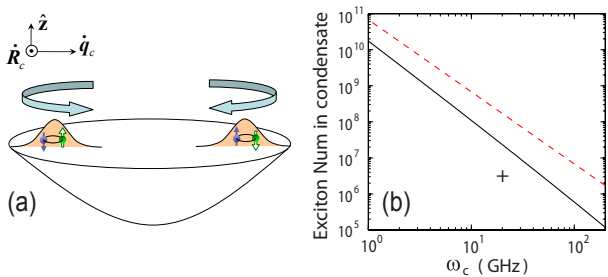}
\caption{(a) Motion of indirect excitons in trap. The Berry
curvature induces a spin-dependent anomalous velocity transverse to
the force from the trap. (b) Critical condensate size as a function
of harmonic trap frequency $\omega_c$. Solid curve shows $N_{\rm
vor}$ above which the condensate is unstable against vortex
formation. Dashed curve denotes $N_{\rm c}$ for BEC-BCS crossover.
For reference, `+' denotes a condensate loaded in a $20$ GHz
trap~\cite{Butov_trap} with a typical density $3 \times
10^{10}$cm$^{-2}$~\cite{Butov_BEC,Butov_coherenceLength}. We have
assumed GaAs coupled quantum well where electron and hole layers are
separated by $d=10$~nm, and the exciton Berry curvature is taken
from the $a_B=10$~nm curve in Fig.~\ref{gaugefield}(b).}
\label{trap}
\end{figure}

{\it Berry Phase Effect on Exciton Condensate.}---In a confining
potential $V(\mathbf{R})$, the anomalous velocity from the Berry
curvature results in a spin-dependent circulation motion for
uncondensed excitons [Fig.~\ref{trap}(a)]. To consider the Berry
phase effect on condensed excitons in a
trap~\cite{Kleppner_BECinTrap}, we first establish the effective
quantum Hamiltonian in presence of the Berry curvature. From
Eq.~(\ref{eom}), it is evident that, with finite $\Omega^{\rm ex}$,
the physical position and momentum of the exciton wavepacket no
longer form a canonical pair. To quantize the non-canonical equation
of motion, the Peierls substitution can be generalized to the
momentum-space gauge field~\cite{Chuu_Dirac}. We find the {\it
canonical} position and momentum variables $\mathbf{R}$ and
$\mathbf{q}$ are related to the physical ones by:
$\mathbf{R}_c=\mathbf{R}+\bm {\mathcal{A}}(\mathbf{q});
\mathbf{q}_c=\mathbf{q}$. The semiclassical energy of the exciton
can then be expressed in terms of the canonical position and
momentum variable as $
\mathcal{E}(\mathbf{R},\mathbf{q})=\mathcal{E}_0(\mathbf{q})+V(\mathbf{R})+\bm{\nabla}
V \cdot \bm{\mathcal{ A}}(\mathbf{q})$. Taking the standard
quantization procedure, we obtain the modified Gross-Pitaevskii
equation with the Berry phase effect
\begin{equation}
\bigg[-\frac{\nabla^2}{2M}+V+U_0 \rho(\mathbf{R}) + \bm{\nabla} V
\cdot \bm{\mathcal{\hat{
A}}}\bigg]\Psi(\mathbf{R})=\mu\Psi(\mathbf{R}), \label{GPE}
\end{equation}
where $\bm{\mathcal{\hat{ A}}} \equiv
\bm{\mathcal{A}}(-i\bm{\nabla})$ is an operator acting on the
condensate wavefunction $\Psi(\mathbf{R})$, and $\rho(\mathbf{R})$
is the exciton density. $U_0=e^2 d/\epsilon$ is the strength of the
repulsive dipole-dipole interaction between indirection
excitons~\cite{Littlewood_Pl_exciton_BEC}. The last term on the
right hand side shows the Berry phase effect. In a harmonic trap
$V(\mathbf{R})=\frac{1}{2}M \omega_c^2 R^2$ with characteristic
length $a_{\rm osc} \equiv (M \omega_c)^{-1/2} \gg a_B$, the Berry
phase term reduces to $\bm{\nabla} V \cdot \bm{\mathcal{\hat{ A}}} =
M \omega_c^2 \Omega_0 \hat{L}_z$ where $\Omega_0 \equiv \Omega^{\rm
ex}(\mathbf{q}=0)$ and $\hat{L}_z$ is the angular momentum operator.

In the multi-component exciton condensate, the Berry phase term is
diagonal in the spin subspace while interconversion between
different spin components is incoherent via the exciton spin
relaxation processes~\cite{Sham_exciton_spin_relax,multiBEC}. We
first analyze how the Berry curvature affect each component.
Obviously, the Berry phase term will lead to a spin-dependent energy
correction to the states with finite angular momentum. To create a
vortex with a single quantized circulation, the cost of energy in
absence of Berry curvature is: $\varepsilon_v=\pi n_0
M^{-1}\ln(0.888 \frac{\lambda}{\xi_0})$~\cite{Pethick_Smith}. $n_0$
and $\xi_0$ are respectively the exciton density and healing length
at the trap center without vortex. $\lambda$ is the spatial
dimension of the exciton condensate which, in the Thomas-Fermi
approximation, is given by $\lambda = a_{\rm osc} \left(U_0 N
M\right)^{1/4}$. The Berry phase term contribute an energy
correction $\Delta \varepsilon_v = M \omega_c^2 \Omega_0
\mathcal{L}_v$ where $\mathcal{L}_v=\pm \frac{1}{2}n_0 \pi
\lambda^2$ is the angular momentum of the vortex
state~\cite{Pethick_Smith}. When $\Omega_0$ and $\mathcal{L}_v$ have
opposite sign, the Berry curvature reduces the energy cost of
creating a vortex in the corresponding spin component of the
condensate. Further, when $\varepsilon_v + \Delta\varepsilon_v <0 $,
a nonrotating condensate component becomes unstable upon forming a
vortex. As $\mathcal{L}_v$ is quadratic while $\varepsilon_v$ is
logarithmic in the condensate size $\lambda$, such instability
occurs when the number of condensed excitons is larger than $N_{\rm
vor} \sim (U_0 M^{3} \omega_c^{2} \Omega_0^{2})^{-1}$.

With the increase of density, the excitonic condensate will cross
from BEC of tightly bound excitons to BCS type momentum-space
electron-hole
pairing~\cite{Littlewood_exciton_BEC_review,Hakioglu_SO_BEC-BCS}. In
the harmonic trap, the border size for such crossover is roughly
$N_{\rm c} \sim (U_0^{-1} M \omega_c^2 a_B^4)^{-1}$, corresponding
to the density $\sim a_B^{-2}$. Thus, for sufficiently large $U_0M$,
$N_{\rm vor} \ll N_{c}$ and the instability is reached well in the
BEC end where the above treatment of condensate is valid. For
typical GaAs coupled quantum wells, our calculations show that
spontaneous vortex formation in BEC phase is expected in tight
confinement which may be realized in electrostatic
traps~\cite{Butov_trap} (see Fig.~\ref{trap}(b)).

{\it Conclusions and Outlooks.}---We have discussed a gauge
structure in exciton wavefunction and the resultant Berry phase
effect on exciton center-of-mass dynamics. A direct consequence is
the spin-dependent topological transport of excitons driven by
mechanical and statistical force, which may add novel spin
functionalities into excitonic circuits~\cite{ExcitonicCircuit}.
Of particular interest is the coupling of the Berry curvature to the
angular momentum of a trapped wavefunction which may lead to
instability of nonrotating BEC above the critical size $N_{\rm
vor}$. The tendency of forming vortex and anti-vortex depends on the
spin species. As instability can be simultaneously reached in each
spin component under equilibrium, an interesting problem of vortex
formation dynamics is posed for future study. If different spin
components are driven into different angular momentum states,
exciton spin relaxation processes can lead to fast depletion of the
condensate. Such dynamics may be experimentally probed from the
angular distribution of
photoluminescence~\cite{Littlewood_Pl_exciton_BEC}.

BEC has also been claimed for exciton-polariton systems
recently~\cite{PolaritonBEC}. It will be interesting to investigate
the Berry phase effect on polariton condensate since polariton can
inherit the Berry curvature from its exciton portion. Because of the
lighter mass of polariton, $N_{\rm vor}$ extrapolated from the above
analysis suggests that Berry phase effect becomes important only at
a much higher density. However, in such case, a dilute BEC may not
be the proper description for the polariton
condensate~\cite{Littlewood_exciton_BEC_review}. Further studies are
needed to have a clear understanding of the Berry phase effects on
other phases of the
condensate~\cite{Hakioglu_SO_BEC-BCS,Ye_phaseDM}.

The work was supported by NSF, DOE, the Welch Foundation, and NSFC.


\begin{thebibliography}{35}
\bibitem{with_TKNN}
With Berry curvature explicitly accounted as a band property, the
semiclassical approach finds Hall conductance as an integral of
Berry curvature over occupied states~\cite{Chang_BerryPhase},
exactly reproducing the linear response result~\cite{TKNN}. The
integral of Berry curvature over the Brillion zone as the quantized
Hall conductance was first recognized in~\cite{TKNN}.
\bibitem{curvature_conduction}
Conduction band Berry curvature plays a less important role due to
the prefactor $\frac{m_e}{M} \ll 1$ when the hole effective mass is
much heavier.
\end{thebibliography}
\end{document}